\documentclass[letterpaper,prl,aps,twocolumn,10pt,longbibliography,superscriptaddress]{revtex4-1}

\usepackage{graphicx,color}
\usepackage{dcolumn}
\usepackage{bm}
\usepackage{mathbbol}
\usepackage{amsmath,amssymb,amsfonts}  
\usepackage[latin9]{inputenc} 

\usepackage[sans]{dsfont}
\usepackage[mathscr]{euscript}

\newcommand{\ket}[1]{|#1\rangle}
\newcommand{\bra}[1]{\langle #1|}

\newcommand{\Hi}{\mathcal{H}}

\newcommand{\Tr}{\mathrm{{Tr}}}
\newcommand{\supp}{\rm{supp}}

\newcommand{\beq}{\begin{equation}}
\newcommand{\eeq}{\end{equation}}
\newcommand{\beqa}{\begin{eqnarray}}
\newcommand{\eeqa}{\end{eqnarray}}
\newcommand{\beqan}{\begin{eqnarray*}}
\newcommand{\eeqan}{\end{eqnarray*}}

\renewcommand{\P}{\mathbb{P}}

\newcommand{\qed}{\hfill $\Box$ \vskip 2ex}

\renewcommand{\span}{{\rm span}}
\newcommand{\rank}{{\rm rank}}

\newcommand{\diag}{{\rm diag}}

\newcommand{\trace}{{\rm Tr}}

\begin{document}

\title{Quantum resources for purification and cooling: \\
fundamental limits and opportunities}


\author{Francesco Ticozzi}
\email{ticozzi@dei.unipd.it}
\affiliation{Dipartimento di Ingegneria dell'Informazione,
Universit\`a di Padova, via Gradenigo 6/B, 35131 Padova, Italy} 
\affiliation{\mbox{Department of Physics and Astronomy, Dartmouth College, 
6127 Wilder Laboratory, Hanover, NH 03755, USA}}

\author{Lorenza Viola}
\email{lorenza.viola@dartmouth.edu}
\affiliation{\mbox{Department of Physics and Astronomy,
Dartmouth College, 6127 Wilder Laboratory, Hanover, NH 03755, USA}}

\date{\today}

\begin{abstract}            
{\bf 
Preparing a quantum system in a pure state is 
ultimately limited by the nature of the system's evolution in the presence of its 
environment and by the initial state of the environment itself. We show that, when 
the system and environment are initially uncorrelated and arbitrary joint
unitary dynamics is allowed, the system may be purified up to a certain (possibly
arbitrarily small) threshold if and only if its environment, either natural or engineered,
contains a ``virtual subsystem'' which has the same dimension and is
in a state with the desired purity.  Beside providing a unified
understanding of quantum purification dynamics in terms of
a ``generalized swap process,'' our results shed light on the significance of a 
no-go theorem for exact ground-state cooling, as well as on the quantum 
resources needed for achieving an intended purification task.}
\end{abstract}

\maketitle


%

Cooling of quantum systems toward their ground state plays a central
role across low-temperature physics and quantum science, by providing
the key to unlock novel phases of matter and quantum behavior -- as
exemplified in settings ranging from laser cooling of atoms and
molecules \cite{Ketterle,bartana-cooling,Cohen-adv} to dynamical
nuclear polarization in solid- and liquid-state nuclear magnetic
resonance, \cite{Abragam,dnp} and cooling of mechanical resonators
\cite{nanomechanical1,Hopkins2003,schwab,cleland,painter}.
From a quantum control standpoint, the task of cooling (or
``refrigeration,'' in the language of quantum thermodynamics
\cite{Kosloff2013}) may be viewed as an instance of dissipative
pure-state preparation, which is in turn closely related to the more
general task of {\em purification} -- namely, the ability to steer the
system from an arbitrary initial state to a final state with higher
purity.  Within quantum information processing (QIP), access to pure
states is presumed in all quantum computation models that can provably
achieve an exponential speed-up over classical ones
\cite{nielsen-chuang,dqc1}, and cold ancilla qubits are critical to
the success of fault-tolerant quantum error correction \cite{QECBook}.
As a result, schemes for cooling and purification are being actively
investigated \cite{Jacobs2013,JacobsWang,mabuchi-hamerly}, and
underlying assumptions and implications formalized with added rigor
\cite{Mahler2011,Paternostro,brumer-cooling,WolfLandauer}.

While in practice {a variety of} system-dependent imperfections and 
technological constraints will inevitably hinder the achievable performance, 
a fundamental question is to determine what ultimate
limitations  may nevertheless exist
on the sole basis of some generic,
{\em system-independent} assumptions on the underlying dynamics.
Specifically, assume that {\em arbitrary} unitary evolution is allowed
on the target system $S$ together with its
environment $E$, starting from arbitrary {\em factorized} initial
conditions.  To what extent does the initial, typically highly-mixed
state of $E$, limit the degree of purity attainable on $S$ {\em in
principle}?  Conversely, if the environment $E$ and its initial state
can be controllably engineered, what are the {\em minimal} resources for
purification (cooling) of $S$ to be guaranteed to a prescribed accuracy?

Our main contribution in this work is the identification of {\em
necessary and sufficient conditions} for exact as well as approximate
purification and ground-state cooling, given the above 
ideal scenario. 
Our starting point is a trivial example: if both $S$ and
$E$ are two-dimensional systems (qubits), purification of 
$S$ is clearly possible in principle only if the initial state of $E$
has a lower entropy, in which case the optimal purification dynamics
simply amounts to swapping the two initial states.  In a general
open-system setting, our strategy is to make precise the intuition
that purity can still only be exchanged but not created between
subsystems, albeit the latter need no longer coincide with the natural
ones.  The relevant notion is provided by the concept of a ``virtual''
subsystem as a {\em factor of a subspace} of a larger state space, as
introduced by Knill {\em et al.}  \cite{viola-generalnoise} in the context
of quantum error correction and extensively used in QIP
\cite{zanardi-virtual,viola-qubit,knill-protected,ticozzi-isometries,viola-IPS,viola-IPSlong}.

Our results complement existing work and advance current understanding
in several ways.  While a no-go theorem for ground-state cooling under
initial system-thermal bath factorization was recently established in
\cite{brumer-cooling}, our analysis further clarifies that such a
no-go strictly applies only to {\em exact} cooling.
Notwithstanding initial factorization, no fundamental limit exists to
{\em arbitrarily accurate} purification and ground-state cooling in
principle, so long as the environment is effectively
infinite-dimensional, and capable of supporting a sufficiently pure
virtual subsystem.  From a quantum-simulation standpoint, this reinforce 
the conclusion that a simulated ancillary environment consisting of a single 
qubit suffices for enacting arbitrary open-system dynamics, so long as it can 
be measured and {\em reset} to a sufficiently pure state \cite{viola-engineering}, 
as recently demonstrated in trapped-ion experiments \cite{barreiro,barreiro2}.
Conceptually, our analysis points to a {\em generalized swap process}
as the unifying physical mechanism through which {\em any}
purification or ground-state cooling dynamics may ensue from joint unitary
evolution, as opposed to {known} special instances limited to small 
dimension and/or a fixed (thermal) initial environment state
\cite{WolfLandauer,brumer-cooling}.  From an open-system
quantum-control perspective, our general picture may be exploited to
design procedures for purification and ground-state cooling via 
environment (or ``reservoir'') engineering, as potentially relevant to a 
growing number of quantum technologies, see
e.g. \cite{poyatos,viola-engineering,ticozzi-steady} and references
therein.  Interestingly, within quantum foundations, our results have
also implications for dynamical reduction models \cite{Ghirardi}: in
order for the ``wave-function collapse'' predicted by the standard von
Neumann postulates to be consistently reproduced by underlying
open-system dynamics, the environment interacting with the system
must, again, {harbor} a sufficiently pure virtual subsystem.

\begin{figure}[t]
\centering
\input{blockdiagram}
\includegraphics[width=9cm]{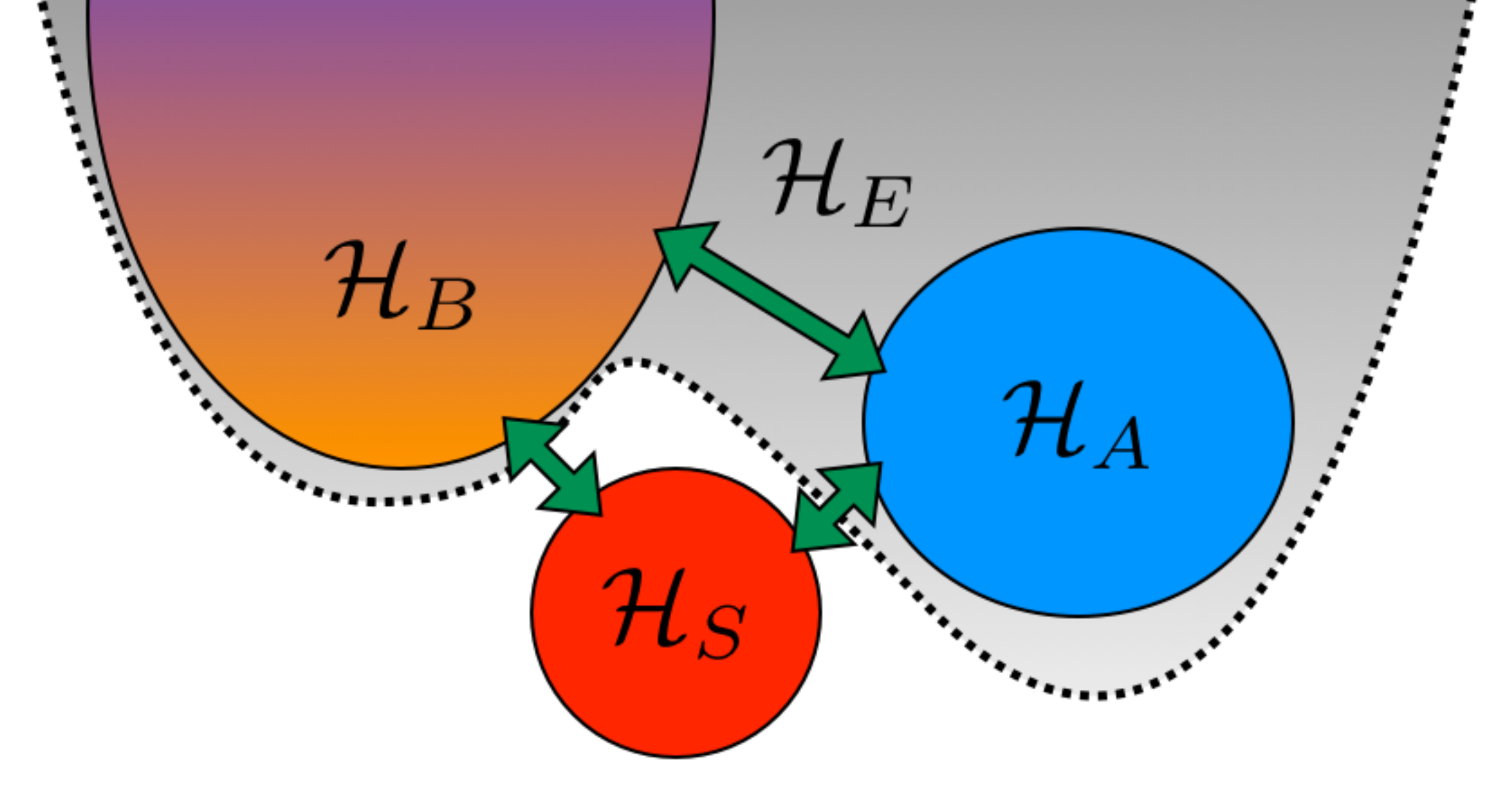} 
\vspace*{-6mm}
\caption{The system of interest, $S$, may be generally coupled to a quantum
bath, $B$, and an engineered auxiliary system, $A$. We 
collectively refer to the pair $(B,A)$ as the {\em environment}. The
initial state on $\Hi_{SE}\equiv \Hi_S \otimes \Hi_E = \Hi_S \otimes
(\Hi_B \otimes \Hi_A)$ is assumed to be {\em fully factorized} with respect
to this partition, i.e.,
$\rho_{SE}=\rho_S\otimes\rho_E=\rho_S\otimes(\rho_B\otimes\rho_A)$.
The joint dynamics is generated by a total Hamiltonian of the form $H
\equiv H_0+H_c(t) = (H_S \otimes {\mathbb I}_E + {\mathbb I}_S \otimes
H_E + H_{SE} ) + H_c(t)$, where the control Hamiltonian $H_c(t) \equiv
\sum_\ell u_\ell(t) H_{c,\ell}$ acts trivially on $B$.  
If dim$(\Hi_{SE})< \infty$, complete propagator
controllability is ensured in the generic case where the Lie algebra
of skew-symmetric operators generated by the control
Hamiltonians $\{ i H_{c,\ell} \}$, together with the natural
``drift'' Hamiltonian $iH_0$, is the whole $\mathfrak{su}(d_S \times d_E)$
\cite{dalessandro-book}. 
If so, there exist some time $T>0$ and control functions $u_\ell(t)$, 
$t \in [0,T]$, that allow to reach any element in $\mathfrak{U}(\Hi_{SE})$ to 
arbitrary precision. For our discussion, it is not essential to specify how the control
actions are enacted.  For instance, if $\Hi_A \simeq {\mathbb C}$,
our setting includes open-loop control of $S$ via a semiclassical
controller \cite{viola-dd,lloyd-coherent}. In this case, $B$ is controlled 
via its interaction with $S$, yet {\em indirect controllability} of $B$ given 
an arbitrary initial state of $S$ still suffices for complete joint controllability, as  
we assume \cite{domenico-indirectcontrollabilty}.   If dim$(\Hi_A)>1$,
dynamics in the presence of a coherent ``quantum controller'' and/or an 
engineered reservoir \cite{lloyd-coherent,JacobsWang} may be accounted for.
In this case, the uncontrollable component $B$ may couple to both 
$S$ and $A$ in general.   }
\label{F1}
\end{figure}
\vspace*{1mm}



\vspace*{3mm}

\noindent
{\bf \large{Results}}

\vspace*{1mm}

\noindent 
{\bf Setting.}  The general setting we consider is depicted in Fig. \ref{F1}.  
The target quantum system $S$, with associated Hilbert space $\Hi_S$ of dimension $d_S$, 
is coupled to a quantum environment $E$, with associated Hilbert space $\Hi_E$ of
dimension $d_E$, which may generally include both a component that is
not directly controllable (a physical ``bath'', $B$) and a fully controllable auxiliary system 
(or ``ancilla'', $A$).   We take $d_E\geq d_S$, so that we may 
decompose $d_E \equiv d_S\,d_F + d_R,$ with $d_F$ being is the integer part 
of $d_E/d_S,$ and $d_R<d_S$ the rest.
While we further assume that $d_S < \infty$ in what follows, we may formally 
extend our results to infinite-dimensional target systems of interest (notably,
quantum oscillators) by imposing a finite-energy constraint. 

A key assumption is that {\em no} correlations are initially present between the
constituents of the joint system, i.e., the initial state is
factorized, $\rho_{SE}=\rho_S\otimes\rho_E$, with $\rho_E$ a
trace-class operator in case $d_E=\infty$.  Other than that, and
unlike \cite{WolfLandauer,brumer-cooling}, {\em no} restriction is placed on
either $\rho_S$ or $\rho_E$ which, in particular, need not be thermal.
We shall denote by $\{ \lambda_j(\rho_E)\}$ the 
eigenvalues of $\rho_E,$ considered with their multiplicity and in 
non-increasing order.

While the inclusion of both a bath and an ancillary system allows for
different physical scenarios to be discussed within the same framework
(see caption), the central mathematical assumption is that suitable
Hamiltonian control is available on $S+E$ together, so that {\em any
unitary operator in ${\mathfrak U}(\Hi_{SE})$ can be obtained at some
time $T$.}  In control-theoretic terms, this is equivalent to assuming complete 
joint propagator controllability \cite{dalessandro-book,ticozzi-intro}. 
Hence, at any given time $T$, the joint evolution of $S+E$ is described by
some $U_{SE}(T)$ that we are free to choose. The conditions for this to
be possible have been extensively investigated within the geometric
control framework. At least if $\Hi_{SE}$ has finite dimension,
complete controllability is generic \cite{dalessandro-book}, albeit
efficient constructive methods for control design are still object of
ongoing research, along with controllability conditions for infinite-dimensional
quantum systems \cite{Rangan,Boscain}.

Starting from factorized initial conditions, the reduced state of the
system after the unitary (controlled) evolution takes place is given by 
\beq
\rho'_S \equiv \trace_E(\rho'_{SE}) = 
\trace_E(U_{SE}\,\rho_S\otimes\rho_E \, U_{SE}^\dag){ .}
\label{eq:reduced}
\eeq {\em Exact purification} of $S$ is attained if $\rho'_S$
is pure {irrespective of the initial state $\rho_S$}, that is, $\rho'_S \equiv 
\ket{\psi}\bra{\psi}$ for some $\ket{\psi} \in 
\Hi_S$, so that $\trace({\rho'_S}^2)=1$. However, this requirement {is}
too strong in practical situations of interest.  We say that {\em
($\varepsilon$-)approximate purification} of $S$ can be attained if 
the state of $S$ may be brought to within distance $\varepsilon$ from 
a pure state {irrespectively of the initial $\rho_S$}, that is, there exists 
$\ket{\psi}\in \Hi_S$ such that:
\beq
d(\rho'_S,\ket{\psi}\bra{\psi}) \leq \varepsilon, \quad 
\forall \rho_S. \label{eq:distance}
\eeq 
\noindent 
Here, $d(X,Y)\equiv \frac{1}{2} \trace(|X-Y|) = \frac{1}{2}
\vert\vert X-Y \vert\vert_1$ is the quantum total-variation distance, which is a natural 
measure of distinguishability between quantum states 
\cite{nielsen-chuang,viola-IPS,ticozzi-isometries}. Exact purification is recovered 
by requesting $\varepsilon=0$. In the following, we shall consider
$0\leq\varepsilon \ll 1$ \cite{Remark}. 

The fact that the joint dynamics $\rho_{SE} \mapsto \rho'_{SE}$ is
unitary is equivalent to the preservation of the spectrum of the joint
density operator at any time. However, one still intuitively expects 
purification of a ``portion'' of the overall system to be possible in an appropriate 
sense, the limitations on what can be achieved stemming from the initial
state of $E$.  Let us first consider a trivial example.

\vspace{0.5mm}

{\em Example 1.--} Suppose that both the target system and the
environment are {a qubit}.  The factorized initial state can then be
parametrized by the maximum eigenvalue of its two components
$\rho_S,\rho_E$, say, $1/2\leq p_S,p_E\leq 1$ respectively, with the
value $1/2$ corresponding to a fully mixed state. That is,
$\rho_{SE}=\diag(p_S,1-p_S)\otimes\diag(p_E,1-p_E)$.  Since for
qubits the von Neumann's entropy ${\cal S}(\rho)$ is completely determined 
by, and is a decreasing function of, the maximum eigenvalue of the state, we can
pursue a direct information-theoretic analysis. Achieving maximal
purification is thus equivalent to achieving the (reduced) state
$\rho'_S$ in Eq. (\ref{eq:reduced}) with minimum entropy with respect
to the choice of $U_{SE}$. Using the standard definitions of joint and
conditional entropy \cite{nielsen-chuang}, we may write
\[{\cal S}( {\rho'_S} )={\cal S}(S)={\cal S}(S,E)-{\cal S}(E|S), \]
\noindent 
where ${\cal S}(S,E)={\cal S}(\rho'_{SE})={\cal S}(\rho_S\otimes
\rho_E),$ and the conditional entropy is maximal when the state is
factorized.  Hence, the maximal purification is attained by either
swapping the states (when $p_E>p_S$), or leaving them as they are
(when $p_E<p_S$). In other words, some purification is possible {\em only if} the 
entropy of the auxiliary qubit is lesser than the one of the system qubit, and 
exact purification can only be achieved if the former is in a pure state to begin with.

\vspace*{0.5mm} 

Despite its simplicity, this example suggests a general strategy to
tackle the purification problem: given a target system to be purified,
if in its environment we may identify a ``subsystem'' of the same
dimension, that is initially in a more pure state, all we need to do
is to swap these two states. Formalizing this idea leads to the 
rigorous conditions we are seeking.

\vspace*{2mm}

\noindent
{\bf Main result: necessary and sufficient conditions for
purification.} In common physical situations, subsystems may be
naturally identified with (distinguishable) quantum particles and/or
degrees of freedom, and their state space directly associated to
different factors of the overall tensor-product Hilbert space. This
view is not, however, sufficiently general to capture all relevant
settings that arise both physically and in the context of QIP
applications.  Within quantum error correction theory, for instance,
``noise-protected'' quantum-information-carrying logical degrees of
freedom are associated with {\em virtual subsystems} that typically do
not correspond with the original qubit subsystems
\cite{knill-QEC,viola-generalnoise,knill-protected}.  
This more general subsystem notion will also be key to our analysis.
Mathematically, a {\em virtual quantum subsystem} $\tilde{S}$ of a
larger system $E$ (the environment in our case) is {associated} with a
tensor factor ${\Hi}_{\tilde{S}}$ of a {\em subspace} of $\Hi_E$
\cite{viola-generalnoise,zanardi-virtual,viola-qubit,knill-protected},
that is, we may write 
\beq \Hi_E = ({\Hi}_{\tilde{S}} \otimes\Hi_F)\oplus \Hi_R,
\label{eq:subs}\eeq
\noindent
for some factor ${\cal H}_F$ and a (generally non-trivial) remainder
space ${\cal H}_R$. System $E$ is said to be initialized in subsystem
$\tilde{S}$ if its state may be decomposed as
$\rho_E= {\rho}_{\tilde{S}} \otimes\rho_F\oplus 0_R,$ where $0_R$ is the
zero operator on $\Hi_R$ and $\rho_F$ a state on $\Hi_F$; in particular, $E$ {\em is initialized 
in a subsystem pure state} if ${\rho}_{\tilde{S}} =\ket{\tilde{\varphi}}\bra{\tilde{\varphi}}$,
for $\ket{\tilde{\varphi}} \in {\Hi}_{\tilde{S}}$
\cite{lidar-initializationfree,ticozzi-isometries,ticozzi-QDS}. While
virtual subsystems are most compactly described in terms of an
operator-algebraic characterization
\cite{viola-generalnoise,zanardi-virtual,viola-IPS,viola-IPSlong}, a
basis with the correct tensor/direct product structure may also be
{straightforwardly} constructed (see Methods). 
We are now ready to state our central result:

\vspace{0.5mm}

{\bf Theorem.} {\em Assume complete unitary controllability and factorized initial conditions 
$\rho_S \otimes \rho_E$ and  on $\Hi_S \otimes \Hi_E$.  Then the following conditions hold:

{\em (i)} For every $\varepsilon > 0$, {\em $\varepsilon$-approximate purification} of $S$ may 
be achieved if there exists a decomposition of $\Hi_E$ as in Eq. \eqref{eq:subs}, 
with ${\Hi}_{\tilde{S}} \simeq \Hi_S$, and a pure-state initialization of $E$ in 
$\tilde S$, $\tilde{\rho}_E =
\ket{\tilde{\varphi}}\bra{\tilde{\varphi}} \otimes \rho_F\oplus 0_R$, such that \beq
\label{condpu} 
d(\rho_E,\tilde{\rho}_E) \leq \varepsilon. \eeq

{\em (ii)} {\em Exact purification} of $S$ $(\varepsilon=0$) may be achieved if and only if the initial state of the 
environment has exactly the form $\rho_E = \ket{\tilde{\varphi}}\bra{\tilde{\varphi}}\otimes \rho_F\oplus 0_R,$ 
for some $\ket{\tilde{\varphi}} \in \Hi_{\tilde{S}}$. 

{\em (iii)} $\varepsilon$-approximate purification is always possible provided that 
$\varepsilon \geq \tilde\varepsilon (\rho_E)$, where  
 \beq 
\tilde\varepsilon (\rho_E) \equiv \tilde\varepsilon =1-\sum_{j=1}^{d_F}\lambda_{j}(\rho_E) \geq 0.
\label{tildeps} \eeq 
$\tilde\varepsilon$-purification 
is optimal whenever $d_R=0.$  In particular, {\em arbitrarily accurate purification} ($\tilde\varepsilon=0$, 
$\varepsilon >0$) is always possible for $d_E=\infty$. } 

\vspace{0.5mm}

Part (i) of the above theorem can be easily proven by considering a unitary operator $W_{SE}$ 
that at some time $T$ swaps the state of $S$ with the one of its isomorphic copy $\tilde{S}$, which
is initially in a pure state $\ket{\tilde{\varphi}}\bra{\tilde{\varphi}}$. 
With the precise definition of $W_{SE}$ being given in the 
Methods section, the basic observation is to note that if $\rho_E$ satisfies
Eq. \eqref{condpu}, then it can be written as 
\beq
\rho_E \equiv \tilde{\rho}_E +\Delta\rho_E,\quad
\frac{1}{2}\trace(|\Delta\rho_E|)\leq\varepsilon.
\label{eq:deviation}
\eeq
By implementing the swap dynamics, it thus follows that
\begin{eqnarray}
\rho'_S & = &\trace_E(W_{SE}\,\rho_S\otimes\rho_E\,
W_{SE}^\dag) \nonumber \\
&= & \trace_E [ \ket{\tilde{\varphi}}\bra{\tilde{\varphi}} \otimes (\rho_{\tilde{S}} \otimes \rho_F\oplus
0_R)+W_{SE}\,\rho_S\otimes\Delta\rho_E \,W_{SE}^\dag] \nonumber \\
&\equiv & \ket{\tilde{\varphi}}\bra{\tilde{\varphi}}  +
\tilde{\cal E}(\rho_S \otimes \Delta\rho_E),
\end{eqnarray}
where $\tilde {\cal E}$ is a trace-preserving completely-positive map
and hence a trace-norm contraction \cite{nielsen-chuang}.  Since,
together with Eq. (\ref{eq:deviation}), this implies that
\[ d(\rho'_S,  \ket{\tilde{\varphi}}\bra{\tilde{\varphi}}   )\leq\varepsilon,
\quad \forall \rho_S, \]
the target system $S$ is $\varepsilon$-purified, as desired. It is immediate to see 
that the same argument also applies if $\varepsilon=0$. In other words, 
the condition of Eq. (\ref{condpu}) is always {\em sufficient} for $\varepsilon$-purification 
with $\varepsilon \geq 0$, independently of the dimension and the initial state of $E$.

Establishing that Eq. (\ref{condpu}) remains {\em necessary} is relatively straightforward 
for exact purification [as in part (ii)], but more subtle in the 
approximate case [part (iii)].  While full proofs are given in the Methods section, 
the gist of the argument showing why 
$\varepsilon$-purification is indeed always possible for 
$\varepsilon \geq \tilde\varepsilon$ may be summarized as follows.
Assume that for an initial state $\rho_{SE}=\rho_S\otimes\rho_E$, 
the desired purification can be attained at some final time $T$. 
Then there exists an orthogonal
projector, say, $\Pi_T=\ket{\psi}\bra{\psi}_S\otimes I_E$, 
such that $\trace(\Pi_T  \rho'_{SE} ) \geq 1-\varepsilon$,
for all $\rho_S.$
If we define a new projector $\Pi_0 \equiv U_{SE}^\dag\Pi_T U_{SE},$ 
this condition clearly also implies that $\trace(\Pi_0 \rho_{SE})\geq 1-\varepsilon.$
This inequality shows that a pure subsystem of dimension $d_S$ 
may be identified to within {distance $\varepsilon$} from the initial {\em joint}
state as well. The tricky part is to establish that this in turn implies the existence 
of an $\tilde\varepsilon$-pure subsystem in the environment {\em alone}. 

In order to do so, we may consider the worst-case scenario, that is, a fully mixed 
(infinite temperature) initial state on $S$, with 
$\rho_{SE} \equiv \tilde{\rho}_{SE}= (1/{d_S}) I_S\otimes\rho_E$.
The idea is to construct a projector of the form $\tilde{\Pi}_0 {\equiv} I_{\tilde{S}} \otimes
{\Pi}_1,$ where ${\Pi}_1$ is a projector on $d_F$ eigenvectors of 
$\rho_E$ with highest eigenvalues, which projects on a subspace, say $\Hi_1 \subsetneq 
\Hi_E$, of the same dimension of $\Pi_T$. This is the best possible strategy whenever $d_R=0$,
and we may show that:
\beq
\trace({\Pi}_1 \rho_E)=\trace(\tilde{\Pi}_0 \tilde{\rho}_{SE})
= 1-\tilde\varepsilon, \quad \tilde\varepsilon \leq \varepsilon.  
\label{eq:subspace}
\eeq
\noindent 
Accordingly, the subspace $\Hi_1$, onto which ${\Pi}_1$ projects,
collects $(1-\tilde\varepsilon)$ of the total probability. The existence of
such a subspace may be shown to be equivalent to the existence of {\em
a virtual subsystem $\tilde{S}$, such that $E$ is 
$\tilde\varepsilon$-close to pure-state initialization in $\tilde{S}$}, as desired. 
  
Our theorem points to an interesting dichotomy between finite- vs.
infinite-dimensional environments.  If $d_E< \infty$,
$\varepsilon$-purification of $S$ may or may not be achievable,
depending on whether the conditions on the spectrum of $\rho_E$
imposed by Eq. (\ref{eq:subspace}) are fulfilled, for arbitrary
$\rho_S$. If $d_E = \infty$, however, then $\tilde\varepsilon=0$ and 
for {\em any} trace-class state of $E$ and {\em any} fixed $\varepsilon >0$, a
sufficiently pure subsystem always exists.
We illustrate how to explicitly construct such a $\varepsilon$-pure
subsystem in the case where the target system is a qubit, as the
generalization to a higher-dimensional system (qudit) is
straightforward.

\begin{figure}[t]
\centering
\input{blockdiagram}
\includegraphics[width=9cm]{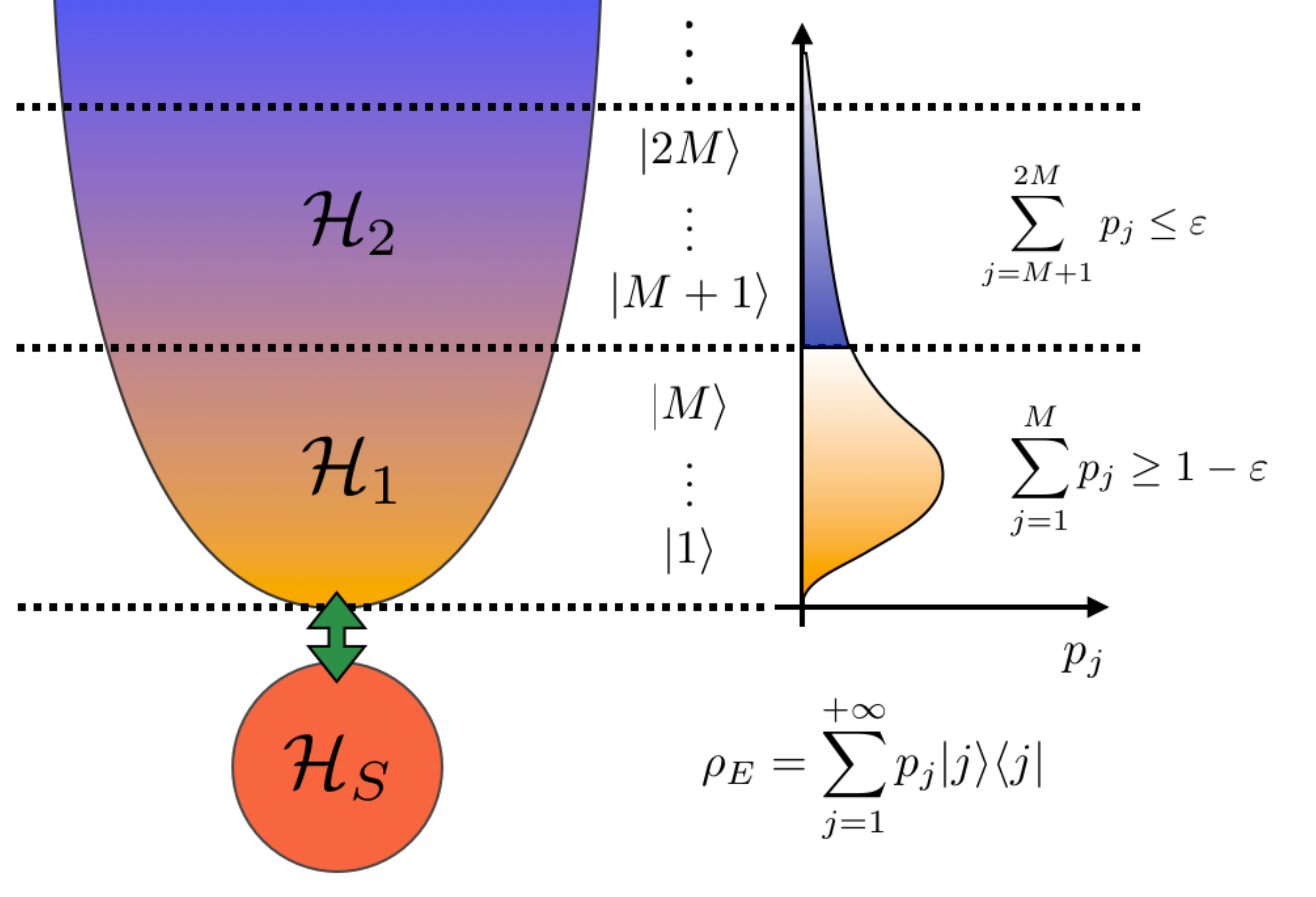} 
\vspace*{-6mm}
\caption{The target system (with $d_S$-dimensional state space $\Hi_S$) 
is coupled to an infinite-dimensional quantum bath
(with state space $\Hi_B$), initially in an arbitrary state
$\rho_B$. To construct a subsystem of $B$ which is 
arbitrarily (yet not perfectly) pure, we identify a finite-dimensional subspace
$\Hi_1$ that collects the first $M$ eigenvectors of $\rho_B$
accounting for $(1-\varepsilon)$ of the total probability. To complete
this virtual subsystem, we only need to identify $(d_S-1)$ orthogonal subspaces
$\Hi_i$, each of dimension $M$. Purification is then attained by swapping
the virtual subsystem's state with the one of the target system. }
\label{F2}
\end{figure}

Let $\rho_E$ be a trace-class environment state, and consider its
spectral representation, {say,} $\rho_E=\sum_{j=1}^\infty p_j
\ket{j}\bra{j}$, $\sum_{j=1}^\infty p_j=1$.  The identification of the
desired $\varepsilon$-pure subsystem may be accomplished by
identifying two orthogonal subspaces $\Hi_1,\Hi_2\subsetneq \Hi_E$ 
each of dimension $M,$ one of which accounts for (at least) $(1-\varepsilon)$
probability.  Since $\rho_E$ is trace class,
hence its spectrum is absolutely summable, for any 
$\varepsilon >0$ there exists an $M$ large enough such that
$\sum_{j>M}p_j<\varepsilon$.  Define $\Hi_1 \equiv
\textrm{span}\left\{\ket{j}\right\}_{j=1,\ldots,M},$ and $\Hi_2$ any
$M$-dimensional subspace orthogonal to $\Hi_1$. From these two
subspaces, we can easily construct a subsystem decomposition 
as in Eq. \eqref{eq:subs}, 
with $\dim(\Hi_{\tilde{S} } )=2$, $\dim(\Hi_F)=M$, such that
the final reduced state $\rho'_S$ is $\varepsilon$-close to a pure
state. The strategy is pictorially illustrated in Fig. \ref{F2}. The
general qudit case can be obtained along the same lines, by
considering $d_S$ copies of the $M$-dimensional subspace, where again
only one accounts for (at least) $(1-\varepsilon)$ of the total probability. 
Similar considerations also apply to typical physical 
scenarios where $d_F \approx d_E/d_S \approx d_E \gg d_S$, 
in which case nearly arbitrary accuracy $\tilde\varepsilon \approx 0$ may 
still be achieved in principle.

Our results show how there is {\em no} fundamental limit to
arbitrarily accurate purification when coupling the target system to
an effectively infinite-dimensional environment. Exact purification, 
on the other hand, would require a sufficiently large number of eigenvalues
of $\rho_E$ to be {\em precisely zero}.  Since this is {\em not} a generic
condition, {in particular it cannot be obeyed if $\rho_E$ is thermal}, 
the no-go theorem of \cite{brumer-cooling} is 
recovered.  With this general conceptual framework in hand, we next
proceed to examine in more detail relevant applications, beginning 
from the special important case of ground-state cooling.


\noindent 
{\bf Ground-state cooling given initial system-bath factorization.}
Consider a setting where, as in Fig. \ref{F2}, the environment
consists of a physical bath ($E\equiv B$), and let $H_S$ denote the
(free) Hamiltonian of the target system $S$, so that
the corresponding initial energy is $\trace(H_S\rho_S).$

Assume first that the minimum eigenvalue $E_{\text{min}}$ of $H_S$
is not degenerate, in which case exact cooling of $S$ to its ground
state entails preparing it in the unique pure state $\ket{\psi_\text{gs}}$ 
corresponding to eigenvalue $E_\text{min}$.  
It is then a straightforward corollary of our theorem that {\em exact
ground-state cooling can be obtained
{\em only if} the environment contains a virtual subsystem of the same
dimension of the target, which is initialized in a pure state.}  Under
the complete joint unitary controllability assumption, however, the
ability to prepare a given pure state also imply the ability to
prepare any pure state in $\Hi_S$. Hence, {\em the existence of a pure
virtual subsytem of the environment is also {\em necessary} for
exact cooling}, fully consistent with the conclusions 
reached in Ref. \cite{brumer-cooling}. 

On the other hand, suppose that only $\varepsilon$-approximate
purification may be achieved in the sense of Eq. (\ref{eq:distance}),
so that the state of $S$ can only be cooled down to within distance
$\varepsilon >0$ from the unique ground state $\ket{\psi_\text{gs}}$
of $H_S$. Then the final energy of the system may be estimated as 
\beqan
\trace[H_S {\rho'_S } ] &=&
\trace[H_S((1-\varepsilon)\ket{\psi_\text{gs}}\bra{\psi_\text{gs}}+\varepsilon
\tau_{\rm ex} ) ] \\ &\leq & (1-\varepsilon) E_\text{min}+ \varepsilon
E_{\text{max}}, \eeqan
\noindent 
where $\tau_{\rm ex}$ and $E_{\text{max}}$ denote some state in the orthogonal
complement to the ground manifold and the maximal eigenvalue of
$H_S$, respectively. Accordingly, approximate ground-state cooling may
be attained with an ``excess'' energy that is upper-bounded by
{$\varepsilon E_\text{max}$.}
We already observed that when $E$ is infinite-dimensional,
$\varepsilon$ can in principle be chosen arbitrarily small, albeit not
zero. Thus, as soon as one allows for approximate yet
arbitrarily good cooling, {\em the no-go theorem can be effectively
evaded} \cite{brumer-cooling}.

If  $E_{\text{min}}$ has degeneracy $d_\text{gs}>1$, 
being able to prepare a pure state still suffices for exact
ground-state cooling, but is no longer needed. 
Sufficient and necessary conditions for
approximate cooling in a degenerate subspace may be derived using
the same reasoning used in establishing necessity of our condition 
for $\varepsilon > \tilde\varepsilon$ -- 
by finding a virtual $d_S$-dimensional subsystem $\tilde{S}$ 
such that $E$ is  $\tilde\varepsilon$-close to initialization in a subspace 
of dimension $d_\text{gs}$ in $\Hi_{\tilde{S}}$.

\vspace*{2mm}

\noindent 
{\bf Arbitrary purification and ground-state cooling with an
engineered qubit reservoir.} 
From an open-system {quantum-}control perspective, our theorem may be 
used to {explicitly characterize what quantum resources may suffice to 
arbitrarily purify/cool the target system}, by coupling it to
a suitably engineered environment ($E\equiv A$). Let us focus on the
simplest yet paradigmatic case in which $S$ is a single qubit and $A$
consists itself of $N$ qubits, so that $\Hi_A \simeq ({\mathbb
C}^2)^{\otimes N}$.

Building on the previous discussion, identifying the desired 
virtual qubit-subsystem entails to split $\Hi_A$ 
into two isomorphic, orthogonal subspaces. For the resulting ``virtual state'' 
to be approximately pure, we further require the probability for $A$ 
to be found in one of such subspaces to be much higher than the 
one for its complementary.
A natural approach is to invoke a ``typical subspace'' argument.
Let each auxiliary qubit be prepared in the same
state, say, $\rho \equiv \diag(q,1-q)$, {$1/2\leq q \leq 1$,}
with respect to a standard 
basis $\{\ket{0},\ket{1}\}$, so that the {joint} initial state 
$\rho_{SA} \equiv \rho_S \otimes \rho^{\otimes N}$.  
As $N$ grows, the state of $A$ will populate
with increasing probability the {\em $\epsilon$-typical subspace}.
Recall that a sequence $x(N)$ of $N$ zeroes and ones, in which each
entry is chosen independently at random with probability
$\P(0)=q$, $\P(1)=1-q,$ is $\epsilon$-typical if \cite{nielsen-chuang}
\[2^{-N \,({\mathcal S}(x)+\epsilon)}\leq \P(x(N))\leq 
2^{-N \,({\mathcal S}(x)-\epsilon)},\]
\noindent 
or, equivalently, its total Shannon entropy is $\epsilon$-close to $N$
times the binary entropy of the single symbol. 
Let ${\mathcal T}(N,\epsilon)$ be the set of $\epsilon$-typical
sequences. In the quantum case, such a set naturally generalizes to
the $\epsilon$-typical subspace: in our qubit setting, the latter
is spanned by those computational basis states that include  
(approximately) $qN$ zeroes and $(1-q)N$ ones: 
\[\Hi_{{\mathcal T}(N,\epsilon)} \equiv \span\{\ket{x(N)}\,|\, x(N)\in 
{\mathcal T}(N,\epsilon)\}.\]

Let now $\Pi_{{\mathcal T}(N,\epsilon)}$ denote the orthogonal projection
onto the $\epsilon$-typical subspace. Then the following asymptotic result
holds (see e.g. Theorem 6.3 in \cite{petz-book}):
{
\begin{equation}
\lim_{N\rightarrow\infty} q_{\text{typ}} (N) \equiv 
\lim_{N\rightarrow\infty}\trace(\Pi_{{\mathcal
T}(N,\epsilon)}\rho^{\otimes N})= 1.
\label{eq:typ}
\end{equation}     }
Furthermore, for any fixed $\epsilon>0$ and a sufficiently large $N$,
the size of the typical subspace satisfies: 
$$ \dim(\Hi_{{\mathcal T}(N,\epsilon)})\leq 2^{N  ({\mathcal
S}(\rho)+\epsilon)}.$$ 
\noindent 
Hence, if $\epsilon$ is sufficiently small, the dimension of the
$\epsilon$-typical subspace becomes {less or equal} than half of the
total space dimension as soon as $N{\cal S}(\rho)< N-1$, or, 
${\cal S}(\rho) < 1- 1/N.$ Therefore, provided that the entropy of 
each of the auxiliary qubits is {strictly} less than one, namely 
$\rho\neq\frac{1}{2}I,$  the typical subspace's dimension will become less 
than half of the dimension of $\Hi_A$ for large enough $N$. If so, we know how 
to explicitly construct a unitary transformation $W_{SA}$ that achieves
(optimal) $\tilde\varepsilon$-purification in principle: it suffices to swap the state of $S$ 
with the state of a virtual qubit system {$\tilde{S}$} that exploits the typical-subspace
structure. We further illustrate this strategy by specializing, again, to ground-state cooling.

\vspace{0.5mm}

{\em Example 2.--}  Assume, similar to Example 1, that the initial state of the target system
$\rho_{S}\equiv \diag(p_S,1-p_S)$,  with respect to the qubit energy basis, say, 
$\{ \ket{\psi_\ell} \} \equiv \{  \ket{\psi_{\text{gs}} }, \ket{\psi_{\text{ex}} } \}$ 
and $p_S <q$. The action of the unitary transformation $W_{SA}$ may be explicitly described 
by introducing a factorized basis $\{ \ket{\psi_\ell} \otimes \ket{j(N)} \}$ on $\Hi_S \otimes \Hi_A$, 
where $\{  \ket{j(N)} \approx  \ket{j_{{\rm typ}}},  \ket{j_{{\rm ntyp}}} \}$ in the large-$N$ limit, 
with $\{ \ket{j_{{\rm typ}}}\} $  and $\{ \ket{j_{{\rm ntyp}}}\} $ denoting orthonormal bases for the 
typical subspace and its orthogonal complement, respectively.
The idea is then to swap $\approx 2^{N{\cal S}(\rho)}$ typical basis states 
which have non-zero probability and are associated to  $\ket{\psi_{\text{ex}} }$, 
with $\approx 2^{N{\cal S}(\rho)}$ non-typical basis states which are in tensor product to 
$\ket{\psi_{\text{gs}} }$ but are associated to low probability. If we compute the final 
energy of the system, by using Eq. (\ref{eq:typ}) we obtain $\trace[H_S \rho'_S  ]
\approx q_{\text{typ}} (N) E_{\text{min}} = (1-\varepsilon) E_{\text{min}} $, with 
arbitrarily small $\tilde\varepsilon$ (hence $\varepsilon$) as $N\rightarrow\infty$, as desired.

\vspace{0.5mm}

Altogether, our results imply that, for a target qubit system, arbitrary accuracy in purification 
and cooling may be achieved through fully coherent (unitary) interaction with sufficiently many 
copies of {\em any auxiliary qubit state which is not the completely mixed one.}
{It is interesting to notice, however, that repeated interactions with an identically prepared 
qubit do not suffice in general: the generalized swap operation needs to simultaneously 
operate on multiple qubits of the engineered environment, pointing to an intrinsic non-Markovian action.  }

\vspace*{2mm}

\noindent 
{\bf Robust pure-state preparation with finite control iterations.} As a final application of our 
framework, our main theorem may be used to understand and characterize the control resources 
involved in a stronger form of purification, whereby the goal is to bring the state of $S$ to a {\em 
predetermined} target pure state $\ket{\psi}_{\rm target} \in \Hi_S$, 
{not necessarily related to the system's ground state}
-- so-called ``global asymptotic stabilization'' in control-theoretic parlance 
\cite{ticozzi-intro,dalessandro-book,ticozzi-QDS,ticozzi-steady}. In particular, the case where 
$\ket{\psi}_{\rm target}$ is an {\em entangled} pure state on a multipartite $n$-qubit target system
provides an important quantum-stabilization benchmark.  
While it is well-known that access to a single {\em resettable} ancillary qubit $A$, 
along with complete unitary control over $S$ and fully coherent ``conditional'' interactions between 
$A$ and $S$, suffices to engineer arbitrary dynamics on $S$ \cite{lloyd-coherent,viola-engineering} 
(and hence achieve the desired stabilization task) in principle, our result sheds light on the 
thermodynamical foundation of this result. With reference to the general setting of Fig. \ref{F1}, 
suppose for simplicity that no uncontrollable bath is coupled to $S$ ($H_{SB}\equiv 0$), 
and that $B$ represents the physical degrees of freedom which enact, possibly together with 
coherent control between $S$ and $A$, the resetting process on $A$. Then, in 
order for stabilization of $S$ to be achievable with arbitrary accuracy $\varepsilon$ starting from 
an {\em arbitrary} environment state $\rho_E \equiv \rho_B \otimes \rho_A$, an effectively infinite-dimensional 
environment is necessary.  Furthermore, {\em exact} pure-state stabilization is only achievable 
provided that $A$ may be perfectly refreshed, which in turn requires $B$ to be perfectly initialized 
in a pure, two-level virtual subsystem. Remarkably, if these conditions are met, an arbitrary $n$-qubit pure state 
$\ket{\psi}_{\rm target}$ may in fact be dissipatively prepared by using a {\em finite} number, $n$, 
of suitably defined control iterations \cite{Baggio2012}. 

Experimentally, controlled dissipation mechanisms are becoming available in a growing number of 
scalable platforms for universal ``digital'' open-system quantum simulators, including trapped-ion 
\cite{barreiro,barreiro2} and superconducting qubit technologies \cite{devoret}. In the above-mentioned 
experiments on $^{40}$Ca$^+$ ions, for instance, the required re-initialization dynamics of the ancilla qubit 
to a reference pure state was realized through a combination of coherent control on $A$, in conjunction with 
optical pumping followed by spontaneous emission. While a number of details are important and require 
careful consideration in practice, conceptually it is this step that ultimately grants access to virtual 
subsystems whose states are sufficiently pure, and can thus be swapped with those of the physical 
degrees of freedom to be purified and/or cooled. 


\vspace*{5mm}

\noindent 
{\bf \large{Discussion}}

\noindent
We have identified sufficient conditions for purification and ground-state cooling of a quantum system 
of interest to be achievable in principle, under the two assumptions of {\em initial system-environment 
factorization} and {\em complete unitary controllability}. 
Such conditions are also necessary in most realistic situations, where the environment is 
much larger than the target system.  While in essence these conditions make rigorous an 
intuition that is both compelling and natural in retrospect -- namely, that purity can only be 
``swapped'' across appropriately defined quantum subsystems -- we have shown how these 
conditions allow to both elucidate fundamental limitations in harnessing open-system dynamics as 
well as identify new opportunities for control engineering.  In particular, our analysis makes it clear 
that {\em arbitrarily {accurate}} purification and/or ground-state cooling is always possible in principle as 
long as the relevant environment is effectively infinite-dimensional, with a no-go result 
\cite{brumer-cooling} only emerging in {the limiting case of zero error}.

From a control-theory standpoint, an interesting direction for further study is to characterize 
what (more stringent) limitations on quantum purification and cooling {may} arise upon relaxing the 
assumption of complete controllability for $S + E$.  
We envision that the existence of a sufficiently pure virtual subsystem in the environment will still 
be a necessary and sufficient condition, albeit identification of the relevant subsystem structure will be 
carried out in this case by exploiting the dynamical-symmetry decomposition associated to the reachable control sub-algebra, in analogy to dynamical error-control strategies and encoded tensoriality 
in QIP \cite{violaDygen,zanardi-ll}.

Lastly, it is interesting to comment on our results in relationship to 
the third law of thermodynamics in its dynamical formulation -- the so-called ``unattainability principle'', 
namely, the impossibility to cool a system to absolute zero temperature in finite time 
\cite{Kosloff2013}.  Throughout our discussion, we have deliberately made no {explicit} 
statement on the time $T$ needed to implement the required generalized swap transformation 
$W_{SE}(T)$.  
For a standard thermodynamic setting where the bath is given, and is initially in a generic 
trace-class state (say, thermal at non-zero temperature), we have showed that arbitrarily small 
cooling error, $\varepsilon >0$, may be achieved only if a sufficiently large subspace of 
the bath can correspondingly account for {\em less than} $\varepsilon$ probability.  
This, in turn, translates into an increasingly complex (energetically ``delocalized'') action of the 
swap transformation $W_{SE}(T)$ to be {implemented}.   Since realistic control Hamiltonians 
are inevitably {\em constrained} (e.g., bounded in amplitude and/or speed, as stressed in 
\cite{Jacobs2013,JacobsWang}), the limit of perfect 
accuracy, $\varepsilon \rightarrow 0$, can only be approached {\em asymptotically} in time, 
$T\rightarrow \infty$. While this supports the validity of the third law under typical conditions, 
it is our hope that our general subsystem-based approach may prove useful to deepen our 
understanding of fundamental performance bounds in more complex thermodynamic scenarios, 
including ``quantum-enhanced'' refrigeration as recently proposed in \cite{Gerardo}.


%

\vspace*{3mm}

\noindent 
{\bf \large{Methods}}
\vspace*{1mm}

\noindent
{\bf Subsystem construction and generalized swap operation.}  Starting
from a general $d$-dimensional Hilbert space $\Hi$, a ``virtual subsystem
structure'' as used in the main text can be identified by constructing a basis 
with the correct tensor/direct sum structure.  The main steps may be summarized 
as follows: \\
\noindent
$\bullet$ Identify a $(d_1\times d_2)$-dimensional subspace $\Hi_{1,2}$, so that 
$\Hi \simeq \Hi_{1,2} \oplus \Hi_R$, where $\Hi_R = \Hi_{1,2}^\perp \equiv \Hi \ominus \Hi_{1,2}$. \\
$\bullet$ Inside {such a} subspace, choose $d_1$ mutually-orthogonal subspaces $\Hi_{j,2}$,
each of dimension $d_2$, so that {we may decompose}
$\Hi_{1,2} \simeq \bigoplus_{j=1}^{d_1} \Hi_{j,2}$. \\
$\bullet$ Pick an orthonormal basis in each of the summands, say, 
$\{\ket{\phi^j}_k , \, k=1,\ldots, d_2\}$, {for $j=1,\ldots, d_1$.} 
We can then establish the following identification: 
\[   \ket{\phi^j}_k=\ket{\phi_j}_S \otimes\ket{\phi_k}_F,  \quad 
\Hi_{1,2} \simeq \Hi_S \otimes \Hi_F,  \]  
and obtain the desired subsystem structure, with dim$(\Hi_S)=d_1$,  dim$(\Hi_F)=d_2$, 
respectively.  

Consider now, specifically, a subsystem structure as given in Eq.~(\ref{eq:subs}) on 
the environment Hilbert space, namely, $\Hi_E = (\Hi_{\tilde{S}}\otimes\Hi_F)\oplus \Hi_R$,
and let $\{\ket{\psi_j}_S\}$, $\{ \ket{{\phi}_k}_{\tilde{S}} \}$, $\{\ket{\xi_\ell}_F\}$, $\{\ket{\chi_m}_R\}$
be orthonormal {(ordered)} bases for $\Hi_S, {\Hi}_{\tilde{S}},\Hi_F,\Hi_R,$ respectively.
We may define the required {\em generalized swap} unitary operator $W_{SE}$ 
through its action on the element of an orthonormal basis. That is, consider the  {(ordered)} basis of
$\Hi_S\otimes\Hi_E$ given by:
\[\{\ket{\psi_j}_S \otimes\ket{{\phi}_k}_{\tilde{S}} \otimes\ket{\xi_\ell}_F\}\cup\{\ket{\psi_j}_S\otimes\ket{\chi_m}_R\},\]
for all $j,k,\ell,m.$ The action of $W_{SE}$ is then defined by:
 \[ \hspace*{-1mm}\left\{ \begin{array}{l}
\hspace*{-1mm}W_{SE} ( \ket{\psi_j}_S\otimes (\ket{{\phi}_k}_{\tilde{S}} \otimes\ket{\xi_\ell}_F ) ) =
\ket{{\psi}_k}_S \otimes \ket{\phi_j}_{\tilde{S}} \otimes\ket{\xi_\ell}_F , \\
\hspace*{-1mm}W_{SE} \left(\ket{\psi_j}_S\oplus\ket{\chi_m}_R \right) =\ket{\psi_j}_S\oplus\ket{\chi_m}_R .
\end{array} \right. \]

\vspace*{2mm}

\noindent 

{\bf Proof the main theorem.}  Assume that, as in the main text, we write $d_E = 
d_S\, d_F + d_R$, with $d_R < d_S \leq d_E$, and let $\rho_{SE}\equiv \rho_S \otimes 
\rho_E$ denote an arbitrary joint initial state on $\Hi_S \otimes \Hi_E$. 

\vspace*{1mm}

{\em Proof of part (ii).}
The fact that the existence of an $\varepsilon$-pure subsystem in the environment 
{\em suffices} for $\varepsilon$-purification  ($\varepsilon \geq 0$) has already been 
proved in the text. We show here that for the case of {\em exact} purification  ($\varepsilon = 0$), 
the existence of a purely-initialized, $d_S$-dimensional subsystem in $\Hi_E$ is indeed also {\em necessary}.

Recall that exact purification is equivalent to the existence of an orthogonal projector, 
$\Pi_T =\ket{\psi}\bra{\psi}_S\otimes I_E$,  such that 
$\trace(\Pi_T \rho'_{SE} ) =1$, 
and that upon defining $\Pi_0 \equiv U_{SE}^\dag \Pi_T U_{SE},$ this also implies that
\beq\label{exsubsys00}
\trace(\Pi_0 \rho_{SE})= \trace(\Pi_0 \rho_{S}\otimes \rho_E )= 1 , \quad  \forall \rho_S.
\eeq
This in particular means that the support of $\rho_S\otimes\rho_E$ is included in the range of $\Pi_0.$
Let us consider $d_S$ specially chosen initial states $\rho_S$, associated to an orthonormal basis 
$\{\ket{\phi_j}_S \}$ of $\Hi_S$, 
that is, 
\[ ( \rho_S\otimes\rho_E )_j = \ket{\phi_j} \bra{\phi_j}_S \otimes\rho_E, \quad j=1, \ldots, d_S. \]
Since their supports are mutually orthogonal, and each of them has dimension $\rank(\rho_E)$, 
it follows that:
\beq \label{sumh} \rank(\Pi_0)\geq \sum_{j=1}^{d_S} \rank (\ket{\phi_S^j}\bra{\phi_S^j}\otimes\rho_E)=d_S\, \rank(\rho_E).\eeq
On the other hand, since rank$(\Pi_T)=d_E$, we also have 
$\rank(\Pi_0)= d_E$. Together with Eq. \eqref{sumh}, this implies:
\[\rank(\rho_E)\leq \frac{d_E}{d_S},\]
and hence, being an integer, $\rank(\rho_E)\leq d_F.$ Call $\Hi_1\equiv\supp(\rho_E)\subset\Hi_E$, 
and construct a $d_S$-dimensional virtual subsystem of $\Hi_E$ as described above. By construction, 
$\rho_E$ is purely initialized in the first elements of the basis associated to the $d_S$-dimensional 
subsystem $\tilde\Hi_S$, leading to the desired conclusion. 

\vspace*{1mm}

{\em Proof of part (iii).} 
%
Let us define the following two quantities [see also Eq. (\ref{tildeps})]: 
\begin{eqnarray*}
\tilde\varepsilon (\rho_E) &\equiv &\tilde\varepsilon =1-\sum_{j=1}^{d_F}\lambda_{j}(\rho_E), \\
\varepsilon_R (\rho_E) &\equiv & \varepsilon_R = \frac{d_R}{d_S} \,\lambda_{d_F+1}(\rho_E).
\end{eqnarray*}
We next proceed to show that:

1. A lower bound $\varepsilon_0$ exists for purification of $S$, independently of the initial state $\rho_S$;  

2. Purification up to $\tilde\varepsilon=\varepsilon_0+\varepsilon_R$  is always possible 
by properly identifying a subsystem in $\Hi_E$ alone and then swapping it with the target.

\vspace*{1mm}

1. {\em  Determining $\varepsilon_0$}.-- We look for necessary conditions on $\varepsilon  > 0$, 
so that $\varepsilon$-purification of $S$ can be attained at time $T$ by some 
joint unitary transformation $U_{SE}$. Again, this means 
that there exists an orthogonal projector,
$\Pi_T=\ket{\psi}\bra{\psi}_S\otimes I_E$, such that 
$ \trace(\Pi_T \rho'_{SE}) \geq 1-\varepsilon$, for all $\rho_S$. 
Upon defining $\Pi_0 \equiv U_{SE}^\dag \Pi_T U_{SE}$ as above, this also 
implies that
\beq\label{exsubsys0}
\trace(\Pi_0 \rho_{SE})=\trace(\Pi_0 \rho_{S} \otimes \rho_E )\geq 1-\varepsilon, \quad \forall \rho_S.
\eeq
Thus, a pure subsystem of dimension $d_S$ may be identified to within 
$\varepsilon$-distance from the {\em joint} initial state as well. While Eq. \eqref{exsubsys0} 
must hold for all $\rho_S,$ in order to determine the desired lower bound we consider 
a worst-case scenario where $\rho_S=(1/d_S) I_S$ and, correspondingly, 
the initial joint state $\rho_{SE}\equiv \tilde{\rho}_{SE}= (1/{d_S}) I_S\otimes\rho_E$.

In fact, consider a basis in which $\tilde{\rho}_{SE}$ is diagonal, ordered in such
a way that its eigenvalues are non-increasing.  The eigenvalues of $\tilde{\rho}_{SE}$ are the
eigenvalues of $\rho_E$, each multiplied by $(1/d_S)$ and repeated $d_S$ times. 
Given that $\Pi_0$ has rank $d_E$, the maximal purification achievable in this case 
correspond to $\Pi_0$ projecting on the first $d_E$ eigenvalues. 
It is then easy to show that:
\beq \label{firstde}\sum_{k=1}^{d_E}\lambda_k(\tilde\rho_{SE})=\sum_{j=1}^{d_F}\lambda_j(\rho_E)+\frac{d_R}{d_S}\lambda_{d_F+1}(\rho_E).\eeq
Since, to guarantee $\varepsilon$-purification, the $d_E$-ranked projector $\Pi_0$ must satisfy 
Eq. \eqref{exsubsys0} in particular for $\rho_{SE}=\tilde{\rho}_{SE}$, Eq. \eqref{firstde} implies the following 
lower bound $\varepsilon_0$:
\beq\varepsilon \geq \varepsilon_0 
\equiv 
\tilde\varepsilon - \varepsilon_R. \eeq
We remark that so far nothing guarantees that $\varepsilon_0$-purification is attainable for 
{\em any} initial state.

\vspace*{1mm}

2. {\em  Attaining $\tilde\varepsilon$-purification}.-- From Eq. \eqref{firstde}, we infer 
that there exists a subspace $\Hi_1$ of $\Hi_E$ {\em alone}, with dimension $d_F,$ that accounts 
for $1-\tilde\varepsilon = 1-\varepsilon_0 -\varepsilon_R$ of the trace of $\rho_E$.
We can thus consider the subspace $\Hi_1 \subsetneq \Hi_E$ that collects {\em only} the one-dimensional 
eigenspaces corresponding to the first $d_F$ eigenvectors of $\rho_{E}$. The last step is to start from 
$\Hi_1$ to construct a virtual subsystem $\tilde{S}$, such that $E$ is $\tilde\varepsilon$-close to 
pure-state initialization in $\tilde{S}$. 

We can in fact identify additional $(d_S-1)$
orthogonal subspaces in $\Hi_E$, say, $\{\Hi_j\}_{j=2}^{d_S}$, all isomorphic to $\Hi_1$
and composed of eigenspaces of $\rho_E$, so that, by following the
general subsystem construction described above, have:
$$\Hi_E  =\Big(\bigoplus_{j=1}^{d_S}\Hi_j\Big)\oplus\Hi_R
\simeq ( {\Hi}_{\tilde{S}} \otimes\Hi_F ) \oplus\Hi_R ,$$ 
\noindent 
where ${\Hi}_{\tilde{S}} \simeq \Hi_S$, dim$(\Hi_F)=d_F$, and 
$\Hi_R =\Hi_E\ominus \bigoplus_{j=1}^{d_S}\Hi_j,$ $\dim(\Hi_R)=d_R.$
Let $\Pi_1$ be the orthogonal projector onto $\Hi_1,$ and 
define $\tilde\Pi_0 \equiv I_S\otimes \Pi_1.$ 
By construction, $\tilde\Pi_0$ has rank $d_S d_F \leq d_E.$  It thus follows that: 
\[  \Tr(\tilde\Pi_0 \tilde\rho_{SE}) =  \Tr(\Pi_1\rho_E) =1-\tilde\varepsilon.   \]
Now notice that with respect to the subsystem decomposition above, we may write 
$\Pi_1=\ket{\tilde\varphi}\bra{\tilde\varphi }\otimes I_F\oplus 0_R$ for some 
$\ket{\tilde\varphi} \in \Hi_{\tilde S},$  and 
\[\rho_E \equiv \tilde{\rho}_E + \Delta \rho_E = 
\ket{\tilde{\varphi}}\bra{\tilde{\varphi}} \otimes \tau_F\oplus 0_R\,  +\,\Delta\rho_E ,\] 
\noindent 
with $\tau_F=\frac{1}{1-\tilde\varepsilon}\,\diag(\lambda_1(\rho_E),\ldots,\lambda_{d_F}(\rho_E)).$ 
Accordingly, with respect to the decomposition $\Hi_E=\Hi_1\oplus\Hi_1^\perp,$ we may write 
$\Delta \rho_E = \Delta\rho_1\oplus\Delta\rho_1^\perp,$ with 
\begin{eqnarray*}
\Delta\rho_1& =& \frac{-\tilde\varepsilon }{1-\tilde\varepsilon}\diag(\lambda_1(\rho_E),\ldots,\lambda_{d_F}(\rho_E)),\\
\Delta\rho_1^\perp &=& \diag(\lambda_{d_F+1}(\rho_E),\ldots,\lambda_{d_E}(\rho_E)).
\end{eqnarray*}
Since these matrices correspond to the positive and negative-semidefinite part of $\Delta\rho_E$, 
it follows that \[\frac{1}{2}\Tr(|\Delta\rho_E|)=\frac{1}{2}[-\Tr(\Delta\rho_1)+\Tr(\Delta\rho_1^\perp)]= \tilde\varepsilon.\]
\noindent 
We may thus conclude that $\rho_E$ admits a $\tilde\varepsilon$-pure subsystem, 
and by using the generalized swapping we can guarantee $\tilde\varepsilon$-purification of 
the target, as claimed.

Note that whenever $d_R=0$, 
we have $\tilde\varepsilon=\varepsilon_0$ and thus our generalized 
swap operator attains the best possible purification.  If, additionally, $d_E=\infty$, 
this also formally corresponds to $d_F=\infty$ hence $\tilde\varepsilon = 0$.   We 
have then explicitly shown in the main text how to achieve purification up to arbitrary 
finite accuracy $\varepsilon >0$. 
\qed
 
%
%

%

\noindent 
{\bf \large{Acknowledgements}}\\
\noindent 
It is a pleasure to thank David Reeb for bringing Ref. \cite{WolfLandauer} to our attention and for 
pointing out a technical problem with an early version of the manuscript, as well as 
Peter Johnson and Alireza Seif for a critical reading of the manuscript. Work at Dartmouth was supported 
in part by the US ARO under contract  No. W911NF-11-1-0068 and the Constance and Walter Burke Special 
Project Fund in {\em Quantum Information Science}. F.T. acknowledges partial support from the QUINTET and QFUTURE projects of the University of Padua.

\end{document}